\documentclass[journal,comsoc]{IEEEtran}

\usepackage[T1]{fontenc}

\ifCLASSINFOpdf
   \usepackage[pdftex]{graphicx}

  \DeclareGraphicsExtensions{.pdf,.jpeg,.png}
\else

\fi

\usepackage{amsmath}

\usepackage[cmintegrals]{newtxmath}

\usepackage[caption=false]{subfig}
\usepackage{algorithm}
\usepackage[noend]{algpseudocode}

\usepackage{graphicx}
\usepackage{latexsym}
\usepackage{amsmath,epsfig}
\usepackage{epstopdf}
\usepackage{caption}
\usepackage{multirow}
\usepackage{booktabs,siunitx}
\usepackage{lscape}
\usepackage{url}
\usepackage{graphicx}
\usepackage{epsfig}
\usepackage{graphicx}
\usepackage{latexsym}
\usepackage{amsmath}
\usepackage{amssymb}
\usepackage{mhsetup}
\usepackage{mathtools}
\usepackage{mathrsfs}
\usepackage{float}
\usepackage{xfrac}
\usepackage{epsfig}
\usepackage[english]{babel}
\usepackage{subfig}

\begin{document}

\title{Robustness Evaluation of the Butterfly Optimization Algorithm on a Control System}

\author{Ramadan~Abdul-Rashid and Basit~Olakunle Alawode \\
	
Electrical Engineering Department, King Fahd University of Petroleum and Minerals, Dhahran, Saudi Arabia \\
Email: {ram.rashid.rr@gmail.com, g201707310@kfupm.edu.sa}
}



\maketitle

\begin{abstract}
In this paper, the Butterfly Optimization Algorithm (BOA) proposed by \cite{arora2018butterfly} is adopted to optimize the parameters of a designed Lead-Lad Controller so as to obtain a stabilized control system. Numerical analysis was carried out for BOA on the control problem and the results are compared to those obtained from the well known Genetic Algorithm (GA) and Differential Evolution (DE) Algorithm.  BOA performs better in terms of eigenvalue analysis but similar to GA and DE in terms of optimizing the minimum damping coefficient for the control system.
\end{abstract}

\IEEEpeerreviewmaketitle

\section{Introduction}
Recent engineering requirements has led to the development of several metaheuristic algorithms that mimic natural phenomenon in order to search for an optimal solution. These algorithms adopt different natural phenomena from the human nervous system used to derive Neural Networks (NN), to the theories of evolutions used to design evolutionary algorithms like the Genetic Algorithm (GA) and the swarming nature of birds, and colonial organisms such as bees used to design the infamous Particle Swarm Optimization algorithm. Owing to the intuitive nature of these algorithms, and their effectiveness in reaching optimal solutions even in non-linear, complex problems, more algorithms continue to be developed based on unexplored natural phenomena and behavior of living organisms. In this paper, we adopt the novel optimization paradigm proposed by Arora \textit{et al} \cite{arora2018butterfly} that impersonates the sustenance searching conduct of butterflies named the Butterfly Optimization Algorithm (BOA). This algorithm been relatively new presents promising results on on a set of 30 benchmark test functions and some classical \textit{mechanical} engineering problems. Despite it performance, to the best of our knowledge, this new paradigm of optimization has not been used to optimize a controller design in a typical control problem. This work seeks to evaluate the robustness of the BOA procedure on a Lead-Lag Controller (LLC) design in other to achieve a desirable frequency response. 

The rest of the paper is organized as follows, Section II expounds on the method of operation and relevant system parameters of the BOA. Section III defines the control system used for the evaluation and formulate the optimization problem. Section IV presents the simulation evaluation of BOA against GA and PSO for the design of an optimal LLC controller. The paper is then concluded in Section V.

\section{Butterfly Optimization Algorithm}
 To comprehend the BOA, some natural certainties and step by step instructions to display them in BOA are examined in following subsections.
\subsection{Butterfly}
In view of logical perceptions, it is discovered that butterflies have an extremely precise feeling of finding the wellspring of scent. Moreover, they can isolate diverse scents and sense their forces \cite{wyatt2003pheromones}. Butterflies are look specialists of BOA to perform streamlining. A butterfly will create aroma with some power which is connected with its wellness, i.e., as a butterfly moves from one area to another, its wellness will fluctuate likewise.

The scent will engender over separation and different butterflies can sense it and this is the means by which the butterflies can share its own data with different butterflies and shape an aggregate social information arrange. At the point when a butterfly can detect aroma from some other butterfly, it will advance toward it and this stage is named as worldwide hunt in the proposed calculation. In another situation, when a butterfly can't sense scent from the encompassing surrounding, at that point it will move haphazardly and this stage is named as neighborhood look in the proposed calculation. In this paper, terms smell and aroma are utilized interchangeably.

\subsection{Fragrance}
In BOA, every aroma has its own special fragrance and individual contact. It is one of the fundamental attributes that recognizes BOA from different metaheuristics. With the end goal to see how aroma is computed in BOA, first we have to get how a methodology like smell, sound, light, temperature, and so on is handled by a stimulus. The entire idea of detecting and handling the methodology depends on three imperative terms viz. sensory modality ($c$), stimulus intensity ($I$) and power exponent ($a$). 

In sensory modality, sensory intends to gauge the type of vitality and process it in comparative ways and methodology alludes to the crude info utilized by the sensors. Presently extraordinary modalities can be smell, sound, light, temperature and in BOA, methodology is scent. $I$ is the extent of the physical/genuine improvement. In BOA, $I$ is associated with the wellness of the butterfly/arrangement. This implies when a butterfly is discharging a more noteworthy measure of scent, alternate butterflies in that encompassing can detect it and gets pulled in toward it. Power is the type to which force is raised. The parameter $a$ takes into consideration customary articulation, direct reaction what's more, reaction pressure. Reaction development is the point at which $I$ builds, the scent ( $f$ ) expands more rapidly than $I$. Reaction pressure is the point at which $I$ builds, f increments more gradually than I. Straight reaction is the point at which $I$ builds, $f$ increments relatively \cite{baird1978fundamentals}. Researchers have led a few investigations on creepy crawlies, creatures and people for greatness estimation and they have inferred that occasionally as the upgrade gets more grounded, bugs turn out to be progressively less touchy to the boost changes \cite{zwislocki2009sensory}. So in BOA, to evaluate the extent of $I$, reaction pressure is utilized.

The common wonder of butterflies depends on two essential issues: the variety of $I$ and definition of $f$. For straightforwardness, $I$ of a butterfly is related with the encoded target work. In any case, $f$ is relative, i.e., it ought to be detected by different butterflies. As indicated by Steven's capacity law \cite{zwislocki2009sensory} with the end goal to separate smell from other modalities, $c$ is utilized. Presently, as the butterfly with less $I$ moves toward butterfly with more I, f builds more rapidly than $I$. So we ought to enable f to fluctuate with a level of ingestion which is accomplished by the power example parameter $a$. Utilizing these ideas, in BOA, the scent is figured as a component of the physical force of improvement as follows:

\begin{equation}
f = c \times I^{a}
\end{equation}

where $f$ is the perceived magnitude of the fragrance, i.e.,
how stronger the fragrance is perceived by other butterflies,
$c$ is the sensory modality, $I$ is the stimulus intensity and a is
the power exponent dependent on modality, which accounts
the varying degree of absorption. For most of the cases in our
evaluation, we can take $a$ and $c$ in the range [0, 1]. The
parameter $a$ is the power exponent dependent on modality
(fragrance in our case) which means it characterizes the changes in absorption. In one extreme, $a = 1$, this means there is no absorption of fragrance, i.e., the measure of aroma radiated by a specific butterfly is detected in a similar limit by alternate butterflies. This is proportionate to stating that aroma is proliferated in a glorified domain. Consequently, a butterfly emanating scent can be detected from anyplace in the space. Subsequently, a solitary (normally worldwide) ideal can be come to effortlessly. Then again, if $a = 0$, it implies that the scent produced by any butterfly can't be detected by alternate butterflies by any stretch of the imagination. Along these lines, the parameter a controls the conduct of the algorithm.

Another important parameter is
$c$ which is also crucial parameter in determining the speed
of convergence and how the BOA algorithm behaves. Theoretically $c \in [0, \infty] $ but practically it is determined by the
characteristic of the system to be optimized. The values of
$a$ and $c$ crucially affect the convergence speed of the algorithm.On account of expansion issue, the force can be relative to the goal work. Different structures of power can be characterized comparably to the wellness work in firefly calculation \cite{yang2010firefly}, hereditary calculations or the bacterial scavenging calculation. 

\subsection{Movement of Butterflies}
To exhibit above dialogs as far as a hunt calculation, the above attributes of butterflies are  idealized as
follows:

\begin{enumerate}
	\item  All butterflies are supposed to emit some fragrance which
	enables the butterflies to attract each other.
	\item  Every butterfly will move randomly or toward the best
	butterfly emitting more fragrance.
	\item The stimulus intensity of a butterfly is affected or determined by the landscape of the objective function.
\end{enumerate}

There are three phases in BOA:(1) Initialization phase,
(2) Iteration phase and (3) Final phase. In each run of BOA,
first the initialization phase is executed, then searching is
performed in an iterative manner and in the last phase, the
algorithm is terminated finally when the best solution is
found. In the initialization phase, the algorithm defines the
objective function and its solution space. The values for the
parameters used in BOA are also assigned. After setting the
values, the algorithm proceeds to create an initial population
of butterflies for optimization. As the total number of butterflies remains unchanged during the simulation of BOA, a
fixed size memory is allocated to store their information. The
positions of butterflies are randomly generated in the search
space, with their fragrance and fitness values calculated and
stored. This finishes the initialization phase and the algorithm
starts the iteration phase, which performs the search with the
artificial butterflies created.

The second phase of the algorithm, i.e., iteration phase,
a number of iterations are performed by the algorithm. In
each iteration, all butterflies in solution space move to new
positions and then their fitness values are evaluated. The algorithm first calculates the fitness values of all the butterflies on
different positions in the solution space. Then these butterflies will generate fragrance at their positions using Eq. (1).
There are two key steps in the algorithm, i.e., global search
phase and local search phase. In global search phase, the
butterfly takes a step toward the fittest butterfly/solution $g^{\textasteriskcentered}$
which can be represented using Eq.(2)

\begin{equation}
x_i^{t+1} = x_i^{t} + (r^2 \times g^{\textasteriskcentered} - x_i^{t}) \times f_i
\end{equation}

where $x_i^{t}$ is the solution vector $x_i$ for $ith$ butterfly in iteration
number $t$. Here, $g^{\textasteriskcentered}$ represents the current best solution found
among all the solutions in current iteration. Fragrance of $ith$
butterfly is represented by $f_i$ and $r$ is a random number in
[0, 1].
Local search stage of the algorithm can be given by:

\begin{equation}
x_i^{t+1} = x_i^{t} + (r^2 \times x_j^{t} - x_k^{t}) \times f_i
\end{equation}

where $x_i^{t}$ and $x_k^{t}$ are $j$th and $k$th butterflies from the solution
space. If $x_i^{t}$ and $x_k^{t}$ t belongs to the same swarm and $r \in [0,1]$ is a
random number then Eq. (3) becomes a local random walk.

Search for nourishment and mating accomplice by butterflies can happen at both nearby and worldwide scale. Considering physical nearness and different components like rain, wind, and so on., search for nourishment can have a critical division $p$ in a general mating accomplice or sustenance looking exercises of butterflies. So a switch likelihood p is utilized in BOA to switch between normal worldwide hunt to concentrated nearby inquiry. Till the halting criteria isn't coordinated, the cycle stage is proceeded. The halting criteria can be characterized in distinctive ways like greatest CPU time utilized, most extreme cycle number achieved, the most extreme number of emphases with no enhancement, a specific estimation of mistake rate is come to or some other proper  criteria. At the point when the emphasis stage is finished up, the calculation yields the best arrangement found with its best
solution found with its best fitness. 

\subsection{Generalized Butterfly Optimization Algorithm}
The complete flowchart of BOA is also as shown in figure 1 and the above-mentioned three steps that make up the complete algorithm of BOA \cite{arora2018butterfly} and its pseudo code is explained in Algorithm 1. 

\begin{figure}[htbp!]
	\centering
	\includegraphics[width=0.45\textwidth]{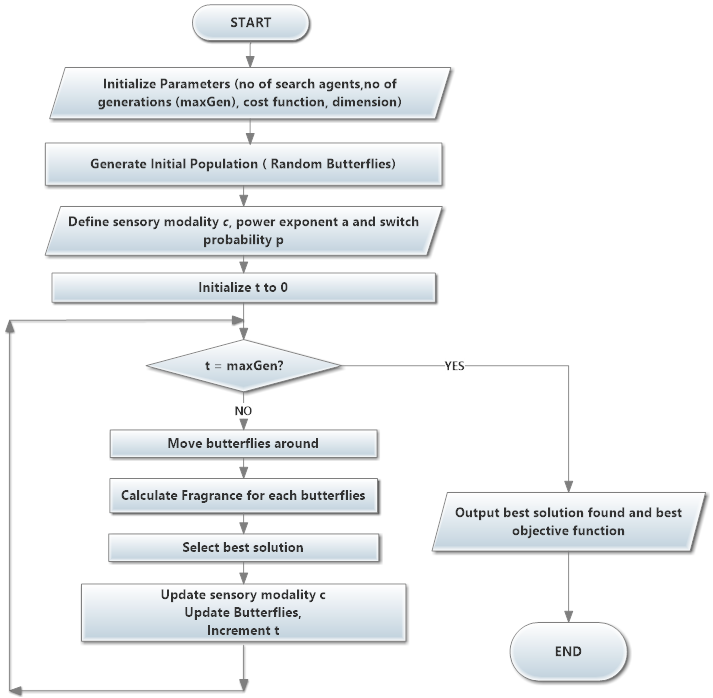}
	\caption[]{BOA Flowchart}
	\label{fig:flowchart}
\end{figure}

\begin{figure}[htbp!]
	\centering
	\includegraphics[width=0.45\textwidth]{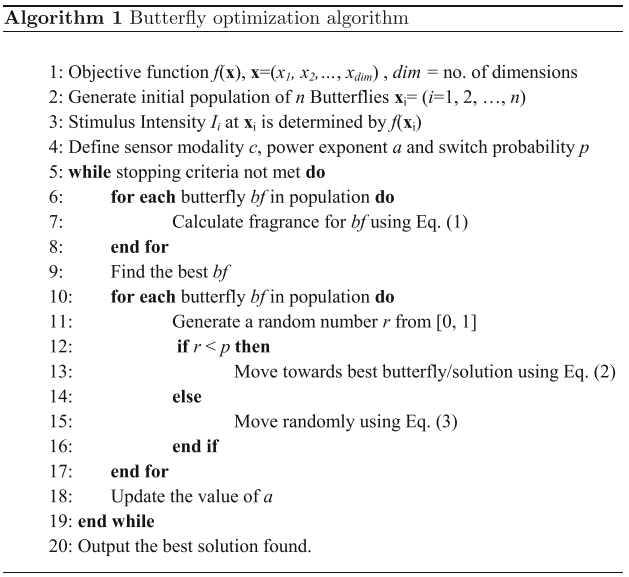}
	\label{fig:alg1}
\end{figure}

\section{Control Problem Formulation}
To test the robustness of the Butterfly Optimization Algorithm (BOA) against other evolutionary algorithms, we formulate a control problem to be used for evaluating BOA. Consider a linearized control system modeled in the state-space form as:

\begin{equation}
\dot{x} = A_o x + B u 
\end{equation}
 Where $A_o \in \mathbb{R}^{n\times n}$, $B \in \mathbb{R}^{n\times 1}$, $x = [x_1(t) \ x_2(t) \ \hdots \ x_n(t)]^T$ and $\dot{x} = [\dot{x}_1(t) \ \dot{x}_2(t) \ \hdots \ \dot{x}_n(t)]^T$

Assuming the control system is unstable with undesirable frequency response, i.e., $|\lambda (A_o)| > 1$. To address this problem, a Lead-Lag compensator (LLC) is designed to mitigate oscillations in the frequency response and also optimize the damping coefficient. The control parameters of an LLC are the controller gain, $K_c$, and time constants, $T_1$ and $T_2$.

The closed loop control system can thus be written as:

\begin{equation}
\dot{x} = A_c x + B u 
\end{equation}
and can be transformed into,

\begin{equation}
\dot{z} = \bar{A}_c z 
\end{equation}

 Where: \\
 $\bar{A}_c \in \mathbb{R}^{m\times m}, \ m = n+2 $, 
 $z = [x_1(t) \ x_2(t) \ \hdots \ x_n(t) \ x_{n+1}(t) \ u]^T$ and 
 $\dot{z} = [\dot{x}_1(t) \ \dot{x}_2(t) \ \hdots \ \dot{x}_n(t) \ \dot{x}_{n+1}(t) \ \dot{u}]^T$
\\

\noindent \textbf{Definition 1}: The damping coefficient, $\zeta$ which describes how fast the system response oscillations decay with time can be related to the complex eigenvalues \cite{abdulkhader2018fractional}, $\lambda_i(\bar{A}_c), \ i = 1, 2, \hdots , m$ of the closed loop system matrix, $A_c$ as:

\begin{equation}
\lambda_i = -\sigma + j\omega_d = -\zeta \omega_{n} + j\omega_{n} \sqrt{1-\zeta^2}
\end{equation}

Where: $\zeta$ is the damping ratio and $\omega_{n}$ is the natural frequency assume to be unity.

From \textbf{Definition 1} and the description of the contro problem, the objective function value associated with the randomly generated solution is found by
calculating the minimum damping ratio ($\zeta_{min}$) which is computed from the minimum  of the closed eigenvalues associated with the values of the controller parameters ($K_c$, $T_1$, and $T_2$). $\zeta_{min}$ can be expressed as:

\begin{equation}
\zeta_{min} = \frac{-\sigma}{\sqrt{\sigma^2 + \omega^2}} = cos\bigg(arctan\Big(\frac{\omega_d}{\sigma}\Big)\bigg) = cos\bigg( arctan \Big(\frac{Im(\lambda_{min})}{-Re(\lambda_{min})}\Big)\bigg)
\end{equation}

We can therefore formulate an optimization problem given as follows:

\begin{equation}
max \ \zeta_{min}
\end{equation}

such that: 
$$K_c^{min} \leq K_c \leq K_c^{max}$$
$$T_1^{min} \leq T_1 \leq T_1^{max}$$
$$T_2^{min} \leq T_2 \leq T_2^{max}$$

\section{Simulation Evaluation}
To evaluate the performance of BOA on the LLC control system, we develop a program to design the optimal controller. This section describes the simulation setup, the system matrices and controller design used. We then carryout  eigenvalue analysis of the system with the optimal controller. The results obtained is compared with those obtained using Genetic Algorithm (GA) and Differential Evolution (DE) algorithm. Genetic Algorithm, one of the most populous evolutionary techniques employs general principles developed in the theory of evolution and genetics to  search for concurrent global solutions to complex optimization problems. The performance of most newly developed evolutionary algorithms are tested against GA \cite{gong2018set}. Differential Evolution (DE) is a widely implemented population based evolutionary optimization technique which is simple, robust, fast and requiring minimal control parameters capable of solving  non-linear and non-differentiable optimization problems \cite{biswas2018optimal}. 

\subsection{Control Problem Parameters}
To conduct our evaluation, we employ an open-loop linearized model system given by:

\begin{equation*}
\begin{bmatrix}
\dot{x_{1}}\\
\dot{x_{2}}\\
\dot{x_{3}}\\
\dot{x_{4}}
\end{bmatrix}=
\begin{bmatrix}
0&377&0&0\\
-0.0587&0&-0.1303&0\\
-0.0899&0&-0.1956&0.1289\\
95.605&0&-816.0862&-20
\end{bmatrix}
\begin{bmatrix}
x_{1}\\
x_{2}\\
x_{3}\\
x_{4}

\end{bmatrix}
.
\end{equation*}
The open loop eigenvalues are the eigenvalues of the matrix A which is:
\begin{equation*}
A=\begin{bmatrix}
0&377&0&0\\
-0.0587&0&-0.1303&0\\
-0.0899&0&-0.1956&0.1289\\
95.605&0&-816.0862&-20
\end{bmatrix}
\end{equation*}

So the open loop eigenvalues are:$ -10.3932\pm 1$ j3.2910 and $0.2954 \pm 1 j4.9577$.
From these eigenvalues obtained, it can be observed that the system is unstable since it contains some of them have  positive real part.\\

\begin{figure}[ht!]
	\centering
	\includegraphics[width= 7.5cm, height= 1.6cm]{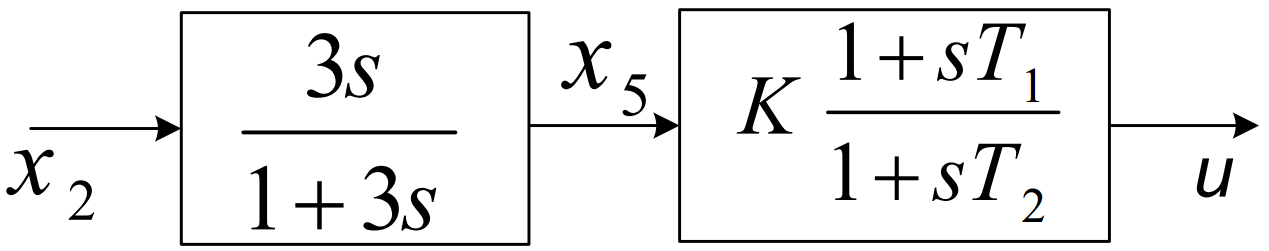}
	\caption{Lead-lag Controller.}
	\label{fig:control_pr}
\end{figure}

The LLC controller with a structure given in Figure \ref{fig:control_pr} is employed for stabilizing the system.

This structure is described by the following equations:

\begin{align}
x_{5}=\dfrac{3s}{1+3s}x_{2}
\end{align}
\begin{align}
x_{5}+3x_{5}s=3x_{2}s
\end{align}
\begin{align}
x_{5}+3\dot{x_{5}}=3\dot{x_{2}}
\end{align}
\begin{align}
\dot{x_{5}}=\dot{x_{2}}-\dfrac{1}{3}x_{5}
\end{align}

To derive the closed loop system, two more states are added to form the transformed model (6) and the system is described by:

\begin{align}
\dot{x_{5}}=-0.0587x_{1}-0.1303_{2}-\dfrac{1}{3}x_{5}
\end{align}
\begin{align}
u=\dfrac{K(1+sT_{1})}{(1+sT_{2})}x_{5}
\end{align}
\begin{align}
u+suT_{2}=Kx_{5}+Ksx_{5}T_{1}
\end{align}
\begin{align}
\dot{u}=\dfrac{KT_{1}}{T_{2}}\dot{x_{5}}+\dfrac{K}{T_{2}}x_{5}-\dfrac{1}{T_{2}}u
\end{align}
\begin{align}
\dot{u}=\dfrac{-0.0587KT_{1}}{T_{2}}x_{1}-\dfrac{0.1303KT_{1}}{T_{2}}x_{3}+\left(\dfrac{K}{T_{2}}-\dfrac{KT_{1}}{3T_{2}} \right)x_5-\dfrac{1}{T_{2}}u
\end{align}
Hence, the new two states will be  $(x_{5}, u)$, therefore, the closed loop state vector will be: $z=〖[x_1,x_2,x_3,x_4,x_5,u]〗^T$\\
Then, the close loop form of state equations of the system is given by:

\begin{equation}
\dot{z}=\bar{A}_{c}z
\end{equation}

Where:
\scriptsize
$$
\bar{A}_c=
\begin{bmatrix}
0&377&0&0&0&0\\
-0.0587&0&-0.1303&0&0&0\\
-0.0899&0&-0.1956&0.1289&0&0\\
95.605&0&-816.0862&-20&0&1000\\
-0.0587&0&-0.1303&0&\dfrac{-1}{3}&0\\
\dfrac{-0.0587KT_{1}}{T_{2}}&0&-\dfrac{0.1303KT_{1}}{T_{2}}&0&\left(\dfrac{K}{T_{2}}-\dfrac{KT_{1}}{3T_{2}} \right)&-\dfrac{1}{T_{2}}
\end{bmatrix}
$$

\normalsize

\subsection{Test Parameters and Setup}
To evaluate the performance of the BOA algorithm against Genetic Algorithm (GA) and Differential Evolution (DE) in solving the control problem in Section IV.A, we develop a programs to design the optimal controller using each scheme. The parameters used for each algorithm in the simulations is given in Table \ref{tab:sim_par}. The limits of the controller parameters adopted for all simulations are: $1.0 \leq K_c \leq 50$, $0.1 \leq T_1 \leq 1.0$, and $0.01 \leq T_2 \leq 0.1$

\begin{table*}[ht!]
	\centering
	\caption{Simulation Parameters for GA, DE and BOA}
	\begin{tabular}{ccc}
		\toprule
		\textbf{GA} & \textbf{DE} & \textbf{BOA} \\
		\midrule
		Mutation Probability = 0.05 & Mutation Probability = 0.05 &  Sensory Modality $c$ = 0.01 \\
		
		Crossover Probability = 0.9 & Crossover Probability = 0.9 &  Switching Probability $p$ = 0.8\\
		
		Crossover coefficient = 0.5 &      &  Power Exponent $a$ = 0.1\\
		
		Population size = 50 & Population size = 50 & Population size = 50 \\
		
		Number of generations = 200 & Number of generations = 200 & Number of generations = 200 \\
		\bottomrule
	\end{tabular}%
	\label{tab:sim_par}%
\end{table*}%

\subsection{Results}
\begin{figure}[ht] 
	\centering    
	\includegraphics[width=0.5\textwidth]{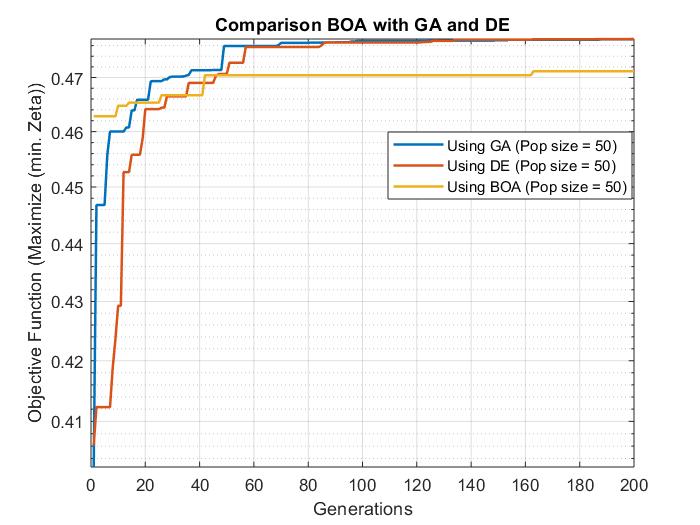}
	\caption[Optimal Objective Function vs. Number of Generations Performance Curves for GA, DE and BOA]{Optimal Objective Function vs. Number of Generations Performance Curves for GA, DE and BOA}
	\label{fig:zeta}
\end{figure}

The code has been run for the same population size and the optimal objective function variation with respect to the the number of generations is shown in Figure \ref{fig:zeta}. It can be noticed that the curves are quite similar to each other which indicates that all algorithms are robust to the same initial population of the solution such that they converge quickly to the optimal values.

\begin{table}[htbp]
	\centering
	\caption{Optimal Controller Parameters}
	\begin{tabular}{crrr}
		\toprule
		Algorithm & \multicolumn{1}{c}{Kc} & \multicolumn{1}{c}{T1} & \multicolumn{1}{c}{T2} \\
		\midrule
		GA   & 18.3998 & 0.2619 & 0.1 \\
		DE   & 18.402 & 0.2618 & 0.1 \\
		BOA  & 18.1352 & 0.2714 & 0.1 \\
		\bottomrule
	\end{tabular}%
	\label{tab:par}%
\end{table}%

It is however observed that the convergence time of BOA is much lower as compared to GA and DE algorithms. Hence BOA outperforms these evolutionary algorithms in term of the time taken to reach the optimal values of $\zeta_{min}$.

Comparing the results of BOA algorithm against GA and DE we observe that eigenvalues of BOA are further left as compared to GA which in turn has eigenvalues further left than DE as shown in Table \ref{tab:eig}. We also observe that BOA achieves a lower constant $K_c$ as compared to the DE and GA. However for the time constant $T_1$, BOA has a higher value as compared to GA and DE but registers the same value of $T2$ as compared to these algorithms. The results for the optimal controller parameters are summarized in Table \ref{tab:par}. On a whole, BOA performs similarly to GA and DE in terms of optimizing the minimum damping coefficient ($\zeta_{min}$). The values for this parameter for each algorithm in the design of the controller is given by Table \ref{tab:zeta} .

\begin{table}[htbp]
	\centering
	\caption{Eigenvalues of Optimal Parameters}
	\begin{tabular}{ccc}
		\toprule
		\multicolumn{3}{c}{Eigenvalues of Optimal Parameters} \\
		\midrule
		GA   & DE   & BOA \\
		\midrule
		-18.2 + 0i & -18.199 + 0i & -18.296 + 0i \\
		-3.032 + 5.5839i & -3.0183 + 5.5576i & -3.2845 + 6.1484i \\
		-3.032 - 5.5839i & -3.0183 - 5.5576i & -3.2845 - 6.1484i \\
		-2.9595 + 5.4499i & -2.9737 + 5.4754i & -2.6591 + 4.9738i \\
		-2.9595 - 5.4499i & -2.9737 - 5.4754i & -2.6591 - 4.9738i \\
		-0.34543 + 0i & -0.34544 + 0i & -0.34519 + 0i \\
		\bottomrule
	\end{tabular}%
	\label{tab:eig}%
\end{table}%

\begin{table}[htbp]
	\centering
	\caption{Optimal Controller Parameters}
	\begin{tabular}{cc}
		\toprule
		Algorithm & Objective Value ($\zeta_{min}$) \\
		\midrule
		GA   & 0.4772 \\
		DE   & 0.4772 \\
		BOA  & 0.4712 \\
		\bottomrule
	\end{tabular}%
	\label{tab:zeta}%
\end{table}%

\section{Conclusion}
In this work, we adopted the relatively new and novel global/local Butterfly Optimization algorithm (BOA) in optimizing the control parameters of a Lead-Lag control system. The general overview and detailed mode of operation of BOA is discussed. A control system with unstable damping coefficient is used to formulate an optimization problem for the design of a optimal controller parameters for a Lead-Lad Controller so as to design a stable control system. On a whole, BOA performs similarly to GA and DE in terms of optimizing the minimum damping coefficient for the cotrol system.

\ifCLASSOPTIONcaptionsoff
  \newpage
\fi

\bibliographystyle{ieeetr}
\bibliography{IEEEabrv,ref}

\end{document}